%% file: main.tex
\documentclass[10pt,conference]{IEEEtran}
\IEEEoverridecommandlockouts
\usepackage{cite}
\usepackage{amsmath,amssymb,amsfonts}
\usepackage{algorithmic}
\usepackage{graphicx}
\usepackage{textcomp}
\usepackage{xcolor}
\usepackage{listings}
\usepackage{booktabs} 
\usepackage{minted}
\usepackage{float}
\usepackage[acronym]{glossaries}
\def\BibTeX{{\rm B\kern-.05em{\sc i\kern-.025em b}\kern-.08em
    T\kern-.1667em\lower.7ex\hbox{E}\kern-.125emX}}
\usepackage{tabularx}
\usepackage{hyperref}
\hypersetup{
    colorlinks=false,
    linkcolor=blue,
    filecolor=magenta,      
    urlcolor=cyan,
    pdfpagemode=FullScreen,
    }

\usepackage[T1]{fontenc}
\usepackage[utf8]{inputenc}
\input{glossaries}
\setminted{fontsize=\small}
\setminted{breaklines}

\usepackage{listings}
\usepackage{xcolor}

\lstdefinestyle{cppstyle}{
    language=C++,
    basicstyle=\ttfamily\small,
    numbers=left,
    numberstyle=\tiny,
    stepnumber=1,
    numbersep=8pt,
    xleftmargin=20pt,
    showstringspaces=false,
    breaklines=true,
    columns=fullflexible,
    keywordstyle=\color{blue},
    commentstyle=\color{green!50!black},
    stringstyle=\color{red!70!black},
    frame=none
}

\title{MQSS Client: Interface for Decoupling Quantum Programming Interfaces}

\author{
\IEEEauthorblockN{Ercüment Kaya}
\IEEEauthorblockA{Technical University of Munich (TUM)\\
Leibniz Supercomputing Centre (LRZ)\\
Garching bei München, Germany \\
ercuement.kaya@lrz.de
}
\and
\IEEEauthorblockN{Muhammad Nufail Farooqi}
\IEEEauthorblockA{
Leibniz Supercomputing Centre (LRZ)\\
Garching bei München, Germany \\
muhammad.farooqi@lrz.de
}
\and
\IEEEauthorblockN{Minh Chung}
\IEEEauthorblockA{
Leibniz Supercomputing Centre (LRZ)\\
Garching bei München, Germany \\
minh.chung@lrz.de
}
\and
\IEEEauthorblockN{Burak Mete}
\IEEEauthorblockA{Technical University of Munich (TUM)\\
Leibniz Supercomputing Centre (LRZ)\\
Garching bei München, Germany \\
burak.mete@lrz.de
}
\and
\IEEEauthorblockN{Martin Schulz}
\IEEEauthorblockA{Technical University of Munich (TUM)\\
Leibniz Supercomputing Centre (LRZ)\\
Garching bei München, Germany \\
schulzm@in.tum.de
}
\and
\IEEEauthorblockN{Jorge Echavarria}
\IEEEauthorblockA{Munich Quantum Valley (MQV)\\
Garching bei München, Germany \\
jorge.echavarria@munich-quantum-valley.de
}
}

\begin{document}

\maketitle
\input{sections/00_abstract}


\glsresetall
\input{sections/01_introduction}

\input{sections/02_background}

\input{sections/03_design_implementation}

\input{sections/04_instantiations}

\input{sections/05_evaluation}
\input{sections/06_conclusion}

\input{sections/07_acknowledgments}

\bibliographystyle{ieeetr} 
\bibliography{bibtex}

\end{document}

%% file: glossaries.tex
\makeglossaries

\newacronym{mqss}{MQSS}{Munich Quantum Software Stack}

\newacronym{hpc}{HPC}{High Performance Computing}

\newacronym{lrz}{LRZ}{Leibniz Supercomputing Centre}

\newacronym{hpcqc}{HPCQC}{High-Performance Computing-Quantum Computing}

\newacronym{qdmi}{QDMI}{Quantum Device Management Interface}

\newacronym{qpi}{QPI}{Quantum Programming Interface}

\newacronym{qc}{QC}{Quantum Computing}

\newacronym{dsl}{DSL}{Domain-Specific Language}

\newacronym{sdk}{SDK}{Software Development Kit}

\newacronym{gpu}{GPU}{Graphics Processing Unit}

\newacronym{fpga}{FPGA}{Field Programmable Gate Array}

\newacronym{qpu}{QPU}{Quantum Processing Unit}

\newacronym{qir}{QIR}{Quantum Intermediate Representation}

\newacronym{ir}{IR}{Intermediate Representation}

\newacronym{qrmi}{QRMI}{Quantum Resource Management Interface}

\newacronym{jit}{JIT}{Just-in-Time}

\newacronym{cc}{CC}{Client Creation}

\newacronym{rq}{RQ}{Resource Query}

\newacronym{jsc}{JSC}{Job Submission and Creation}

\newacronym{eqc}{EQC}{Execution of Quantum Circuit}

\newacronym{res}{RES}{Result Retrieval}

\newacronym{te}{TE}{Total Execution}

%% file: sections/00_abstract.tex
\begin{abstract}
\gls{qc} is an emerging technology that requires customized tools, such as software stacks and programming interfaces.\ 
However, currently, the tools are generally tightly coupled and exhibit limited interoperability.\
This, in particular, affects \gls{hpc} facilities and data centers, which are required to support multiple programming interfaces.\
In this paper, we introduce MQSS Client, an unifying, context-aware access layer and programming library that decouples the programming interfaces and the underlying compilation and runtime stack.\
MQSS Client aims to support all existing programming interfaces by providing abstractions for resources, jobs, and results.\
It provides two access modes to accommodate the varied needs of remote and \gls{hpc} users.\
Thus, interoperability between software stacks and programming interfaces increases.
\end{abstract}

%% file: sections/01_introduction.tex
\section{Introduction}

\gls{qc} is advancing swiftly and promises to revolutionize computation by enabling solutions to problems that would require exponential resources for classical computing systems. Rather than replacing classical infrastructure, \glspl{qpu} are increasingly envisioned as accelerators—similar to \glspl{gpu} and \glspl{fpga}—within existing \gls{hpc} environments.

While remarkable progress has been made in hardware development, such as Google's Willow~\cite{neven-2025} and IQM superconducting quantum processors~\cite{mansfield2025iqmfirstpractice}, these achievements alone are insufficient to unlock the full potential of \gls{qc}.
To achieve the quantum advance truly, the hardware requires compatible and sophisticated programming languages, models, and robust underlying compilation infrastructures.\
The challenge lies not only in hardware limitations but also in the lack of cohesive software ecosystems that allow users to effectively develop, execute, and manage quantum applications across heterogeneous platforms. These components are crucial for effectively exploiting the quantum systems.

A \gls{qpu} in quantum technologies can be implemented using different physical approaches, such as superconducting or photonic systems.\
This diversity affects the usability of the devices.\
For instance, the definition of a \textit{quantum job} may vary between vendors.\
In parallel, this diversity poses challenges for data centers and \gls{hpc} facilities that host multiple distinct quantum devices.\
Hence, unified software stacks that support different devices have become increasingly important.

\begin{figure}[!t]
    \centering
    \includegraphics[width=\columnwidth]{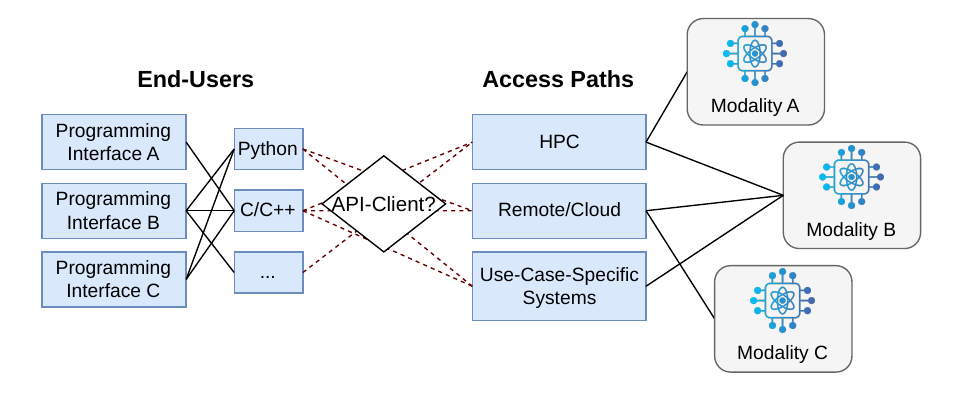}
    \caption{An illustration of multi-programming models and multi-quantum modalities connecting through different access paths, underscoring the motivation for a unified API client.
    }
    \label{fig:mqss-client-motiv}
\end{figure}

The thriving field of \gls{qc} led to the development of a wide range of programming models and \glspl{dsl}.\
Figure~\ref{fig:mqss-client-motiv} illustrates a scenario from a computing center side, where there might be multiple programming interfaces for end-users.\
Users can expect to work across multiple quantum modalities via different access paths and choose the most suitable option based on their expertise and needs.\
Similarly, this diversity poses a challenge for the mentioned facilities, as they are required to support a wide range of \glspl{dsl}.\
\gls{sdk} fragmentation is the central obstacle: each quantum vendor ships its own incompatible toolchain, forcing HPC facilities to maintain parallel, redundant software stacks.

In this paper, we introduce the \textit{MQSS Client}, a unifying, context-aware access layer that addresses the fragmentation of the quantum software stack by decoupling the programming model from the underlying compilation and runtime stack.\
\textit{MQSS Client} is designed to support a broad range of existing programming interfaces via middleware modules named \textit{MQSS Adapters} \cite{mqss_interfaces}.
In particular, \textit{MQSS Client} supports multiple execution environments, including REST-based remote/cloud access, \gls{hpc}, or use-case-specific systems, allowing it to operate seamlessly across diverse deployment scenarios. 
By acting as a bridge between programming interfaces and quantum backends, it enables users to develop and submit quantum jobs in a consistent manner, independent of the underlying system.

In summary, our contributions are as follows:
\begin{itemize}
    \item a unified abstraction layer that decouples programming models from backend-specific infrastructures;
    \item a flexible interface supporting multiple access paradigms across remote/cloud, \gls{hpc} environments, or different use-case-specific systems;
    \item a runtime library for standardized quantum job submission, resource querying, and result retrieval.
\end{itemize}

The remainder of this paper is organized as follows: Section~\ref{sec:background-relatedwork} covers background and related works. In Section~\ref{sec:design-implementation}, we introduce the design and implementation details of \textit{MQSS Client}. Section~\ref{sec:instant-deploy} describes the instantiations and how we deploy the client in our ecosystem. In Section~\ref{sec:exps-evaluation}, we present the experiments and performance evaluation. Finally, we conclude the paper in Section~\ref{sec:conclusion}.

%% file: sections/02_background.tex
\section{Background and Related Work} \label{sec:background-relatedwork}



Quantum programming models and interfaces are essential to interact with quantum hardware.\
While programming models define how computation is performed, interfaces are practical tools to enable creating quantum programs.\
The most well-known quantum programming models are gate-based, pulse-based, and Hamiltonian models.\
The gate-based model defines quantum circuits as a sequence of quantum gates.\
Similarly, the pulse-based model uses a sequence of physical signals.\
On the other hand, the Hamiltonian model describes the computation via the system’s energy function.\
To describe a quantum program at a higher level, we use programming interfaces, such as Qiskit \cite{qiskit} and CUDA Quantum \cite{cuda-q}.

Python is the dominant language in the \gls{qc} domain.\
For instance, Qiskit \cite{qiskit} is a Python-based, open-source \gls{sdk} developed by IBM and widely used in developing quantum applications.\
It provides abstractions to define custom backends and quantum job definitions.\
Another widely used Python-based \gls{sdk} is PennyLane \cite{pennylane}, developed by Xanadu Quantum Technologies.\ 
As an extension to it, Catalyst is a \gls{jit} compiler that compiles \gls{hpcqc} applications as a whole~\cite{Ittah2024}.\
PennyLane also provides an abstraction to create a custom backend.\
To enable the \gls{jit}, an additional custom backend needs to be created by inheriting the corresponding interface from Catalyst's Runtime.

While Python dominates quantum application development, it is unsuitable for HPC environments where low-latency, compiled code is standard.\
C and C++ remain the dominant languages in HPC, making native support for them essential for seamless \gls{hpcqc} integration.\
Therefore, it is necessary to create programming interfaces in these languages to enable seamless \gls{hpcqc} integration.\
For instance, CUDA Quantum \cite{cuda-q} is a language extension that allows users to integrate quantum circuits using C++ and Python.\
It allows the creation of different backends with their abstraction classes.
In parallel, \gls{qpi} is a C-based runtime library solution that allows users to create quantum circuits within \gls{hpc} applications.\
In addition, CQ \cite{cq} is a C-like API, programming model, and specifications for the hybrid application. 

In parallel, quantum software stacks, such as \gls{mqss} \cite{disaggregated-paper, mqss-paper}, are required to support multiple programming models and quantum devices.\
Similarly, Rust-based \gls{qrmi} \cite{bacher2025quantum} is a vendor-agnostic library for submitting quantum jobs and monitoring their behavior by abstracting away the complexity of quantum resources.\ 
On the backend side, \gls{qrmi} provides an abstraction layer that supports multiple quantum devices.\
In addition, \gls{mqss} provides abstraction via \gls{qdmi} \cite{qdmi}.\
Hence, abstraction is a key concept for stripping away the complexity and versatility of the fast-paced development of the quantum ecosystem.

%% file: sections/03_design_implementation.tex
\section{Design and Implementation} \label{sec:design-implementation}


\begin{figure*}
    \centering
    \includegraphics[width=\linewidth]{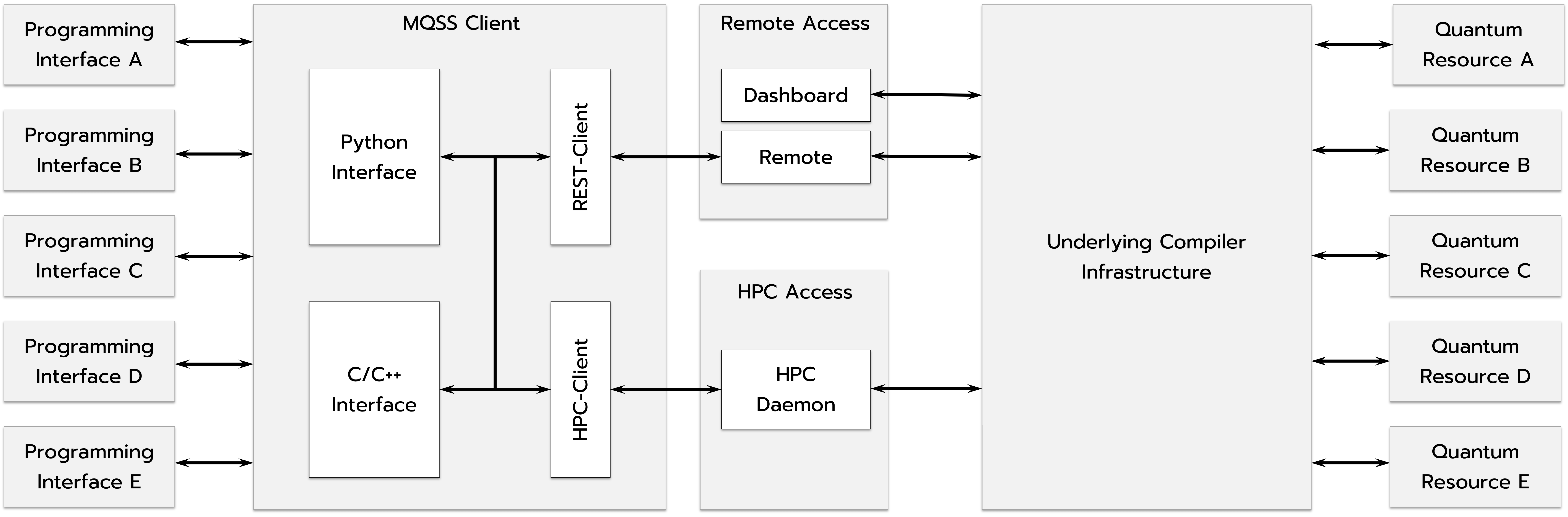}
    \caption{The structure of \gls{mqss} Client: 
    It is designed to be between programming interfaces and the underlying compiler infrastructure.}
    \label{fig:mqss-client}
\end{figure*}

As depicted in Figure~\ref{fig:mqss-client}, the \textit{MQSS Client} is a unifying, context-aware access layer and programming library that decouples the programming interfaces and the underlying compilation and runtime stack.\
We implement the \textit{MQSS Client} using C++17 with Niels Lohmann's JSON \cite{Lohmann_JSON_for_Modern_2025} to handle the JSON data format, \texttt{libcurl} \cite{curl} to establish the remote connection, \texttt{rabbitmq-c} \cite{rabbitmqc} to handle the \gls{hpc} connection.
To create the Python bindings, we utilize \texttt{pybind11} \cite{pybind11}.

\begin{figure*}[!t]
    \centering
    \includegraphics[width=2\columnwidth]{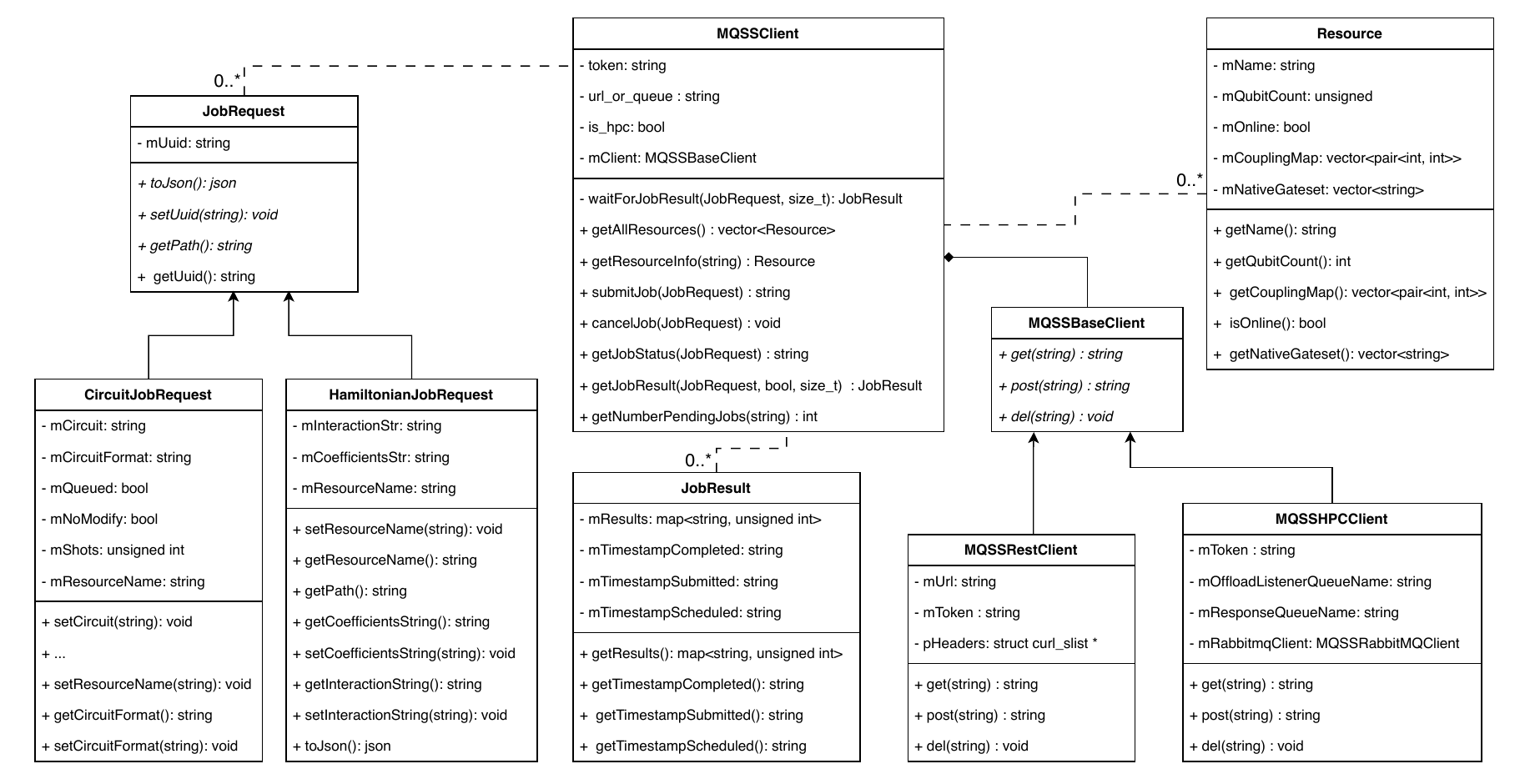}
    \caption{The class diagram of the \textit{MQSS Client} that shows the relationships among the classes and the abstractions}
    \label{fig:mqss-client-classdiagram}
\end{figure*}

Figure \ref{fig:mqss-client-classdiagram} shows the full class 
structure of the \textit{MQSS Client}.\
At the top level, \texttt{MQSSClient} is the single entry point for all user interactions.\
It delegates transport to one of two internal backends ---\texttt{MQSSRestClient} or \texttt{MQSSHPCClient}--- both implementing the 
\texttt{MQSSBaseClient} interface.\ 
This design localizes  transport-specific logic within the backend classes, keeping the public-facing \texttt{MQSSClient} API uniform regardless of  the chosen access mode.\
Outward from \texttt{MQSSClient}, three  domain objects carry the core abstractions: \texttt{JobRequest} defines what to run, \texttt{JobResult} captures what came back, and \texttt{Resource} describes where it ran.


We provide two access modes to accommodate distinct requirements: a REST-based remote access and a RabbitMQ-based \gls{hpc} access.\
The access modes are built on top of the unified layer that encapsulates the job and resource interfaces. \
This distinction allows compilation and runtime stacks to follow different protocols while remaining decoupled. 

The \textbf{REST-based remote access} mode allows users to access quantum resources in a remote location by exposing the resources through standard HTTP endpoints. \
In parallel, the \textbf{RabbitMQ-based \gls{hpc} access} is tailored for performance-critical environments such as data and supercomputing centers. \
While Listing \ref{listing:client-usage} demonstrates the life cycle of the \gls{mqss} Client, the line 4 creates an object based on \gls{hpc} access, in parallel, line 5 demonstrates object creation for remote access.

\begin{listing}[ht]
\begin{lstlisting}[style=cppstyle]
#include <mqss/client.h> 
int main() { 
    ....
    auto client = MQSSClient(token, queue_or_url, isHpc);

    auto resources = client.getAllResources();
    auto resourceToUse = chooseTheResource(resources);
    std::string resourceName = resourceToUse.getName();

    std::string circuit = R"(
        OPENQASM 2.0;
        include "qelib1.inc";
        qreg q[2];
        creg c[2];
        h q[0];
        cx q[0], q[1];
        measure q -> c;)";
    
    CircuitJobRequest circuitJob = CircuitJobRequest(circuit, "qasm", resourceName, shotCount, noModify, isQueued);
    auto circuitJobID = client.submitJob(circuitJob);

    std::string interactionStr = "0 1; 1 2; 0 2; 0 3;";
    std::string coefficientsStr = "0.5 0.1 0.8 1;";
    HamiltonianJobRequest hamiltonianJob = HamiltonianJobRequest(resourceName, interactionStr, coefficientsStr);
    auto hamiltonianJobID = client.submitJob(hamiltonianJob);

    std::unique_ptr<JobResult> hamiltonianResult = client.getJobResult(hamiltonianJob, true, 10);
    std::unique_ptr<JobResult> circuitResult = client.getJobResult(circuitJob, true, 10);
    
}

\end{lstlisting}
\caption{The Life Cycle of the MQSS Client}
\label{listing:client-usage}
\end{listing}

We provide two interfaces within the \gls{mqss} Client to fulfill the requirements of end users and libraries built on top of it: the Job and Resource Interfaces. \
At a high level, a \textit{Job} represents a \textit{quantum task} while a \textit{Resource} represents any quantum computation asset that can execute a quantum task, such as quantum computers, simulators, as well as stacks.

The \textbf{Job interface} encapsulates the logic related to quantum task definition and execution.\ 
The job interface consists of highly customizable \textit{Job Request} and \textit{Job Result} classes.\ 
A Job Result includes the outcome of the quantum task and execution metrics, such as completion and submission times.\
Conversely, a Job Request defines the quantum task to be offloaded to the underlying stack.

    
    

%
%

We define two job requests to accommodate the use of different quantum-computation assets: \textit{Circuit} and \textit{Hamiltonian}.\
A circuit job request describes a gate-based quantum program, including the number of shots and the resources to be used during execution. \
On the other hand, a Hamiltonian job request describes a quantum task that determines the total energy of a quantum system with given interaction terms and coefficients.\
Lines 11 to 21 in Listing~\ref{listing:client-usage} show the creation and submission of a circuit job request, followed by a Hamiltonian job request at lines 23 to 26 in the same Listing.

The \gls{mqss} Client can query the results of a job using the \textit{getJobResult} function, where the first parameter is the job to be queried, followed by a boolean value indicating whether to wait for the results, and an integer value for the timeout. \
In that case, if the results are not produced or not ready, the returned value is NULL. \
Otherwise, the function returns a unique pointer to a \textit{Job Result}.\
It contains the histogram of the results, along with metrics such as submission time and completion time.

The \textbf{Resource Interface} dynamically abstracts the characteristics and capabilities of the available quantum computation assets. \
It exposes methods such as \textit{getCouplingMap}, \textit{getQubitCount}, and \textit{getNativeGateset} that provide information about the resources' capabilities. \
Lines 7 to 9 in Listing \ref{listing:client-usage} show querying the resources and receiving the properties of a resource.

%
%
%


%% file: sections/04_instantiations.tex
\section{Instantiations and Deployment} \label{sec:instant-deploy}

This section describes how the \gls{mqss} Client is instantiated and deployed in different programming languages, programming models, and configurations.\ 
Its low-level implementation and bindings allow us to exploit it as a standalone client library as well as a backend service for the external applications. \
This flexibility improves the re-usability of the MQSS Client.

The \gls{mqss} Client is deployed at one of Europe's leading supercomputing centers for \gls{hpcqc}, and used by a diverse user base with different programming languages and models—from beginner users who use it in remote mode to advanced users who use it in tightly-coupled hybrid applications.

We divide the instantiations of \gls{mqss} Client into two: \textit{Standalone Client Library} and \textit{Backend Adapters}, and Table~\ref{tab:instantiations} summarizes the instantiations based on language, access mode, and the job type.

\begin{table}[]

\resizebox{\linewidth}{!}{
\begin{tabular}{llcc}
\toprule
\textbf{Instantiation} & \textbf{Language} & \textbf{Access Mode}& \textbf{Job Types} \\
\midrule
Standalone Library        & C++, Python & Both & Circuit, Hamiltonian  \\
\gls*{qdmi} Device         & C++         & Remote & Circuit\\
\gls{qpi} Backend         & C           & HPC & Circuit\\
CUDA-Q Backend            & C++         & Both & Circuit\\
Qiskit Backend            & Python      & Both & Circuit\\
PennyLane Backend         & Python      & Both & Circuit\\
Benchmarking Framework    & Python      & Remote & Circuit\\
\bottomrule\\
\end{tabular}
}
\caption{Overview of \gls{mqss} Client Instantiations}

\label{tab:instantiations}
\end{table}
 
\subsection{Instantiation as a Standalone Client Library}

In this instantiation, the \gls{mqss} client is embedded within an \gls{hpcqc} application that is directly exposed to end users.\
The users are responsible for orchestrating interactions with the library.\
This instantiation is most suitable for the \gls{hpcqc} application with static quantum circuits.

As given earlier, Listing \ref{listing:client-usage} presents an example on using the \gls{mqss} Client as a standalone library.\
Thanks to the \gls{mqss} Client's Python bindings, it can be used in Python applications.\

\subsection{Instantiations as Backend Adapters}

In this instantiation, the \gls{mqss} Client is wrapped around the existing backend configurations to provide support for programming models and languages.\
The configurations support both REST-based remote and RabbitMQ-based \gls{hpc} access modes, determined by the environment in which they are initiated.\

\subsubsection{Stack-as-device \gls{qdmi} Device}\hfill\break
\gls{qdmi}, one of the core components of the \gls{mqss}, is a C-based API that encapsulates job submission, capability queries, and the life cycle of a quantum system.\
Any implementation of \gls{qdmi} specifications referred as \gls{qdmi} Device \footnote{They can be found here: https://github.com/Munich-Quantum-Software-Stack/MQSS-QDMI-Devices-Suite}.\
The majority of the \gls{qdmi} devices correspond to a quantum asset.\
However, the \texttt{Stack-as-device} concept enables us to implement the \gls{qdmi} specification as a software stack.\
In this way, various software stacks may exploit the quantum resources of other facilities and software stacks. 

We implement the specifications of the \gls{qdmi}, using \gls{mqss} Client to create a \gls{qdmi} device for \gls{lrz}.\
The most critical components of the implementation are outlined below:
\begin{enumerate}
    \item The \textbf{QDMI\_device\_initialize} function initiate the device.
    \item The \textbf{QDMI\_device\_session\_init} function initiate the session based on the given parameters, such as \textit{base URL} and \textit{token}, that set using \textbf{QDMI\_device\_session\_set\_parameter} function.
    \item The \textbf{QDMI\_device\_job\_submit} function submits a circuit job, which is created and configured by using the \textbf{QDMI\_device\_session\_create\_device\_job} and \textbf{QDMI\_device\_job\_set\_parameter} functions, respectively. 
\end{enumerate}

\subsubsection{C-based \gls{qpi} Backend}\hfill\break
As mentioned earlier, \gls{qpi} is a lightweight, C-based programming library that allows embedding quantum circuits in \gls{hpc} applications.

To connect \gls{qpi} to \gls{mqss} Client, we embedded it into the runtime using C bindings.\
Hereby, users can submit a quantum job in a location and \gls{hpc} environment without any modifications in the code.

\subsubsection{C++ based CUDA-Q Backend}\hfill\break
CUDA-Q \cite{cuda-q} allows the implementation of custom backends using the provided abstractions.\
Its abstractions include, but are not limited to \textit{QPU} class which executes the quantum kernels, and \textit{Executor} which executes compiled quantum codes.\

To connect CUDA-Q to \gls{mqss} Client, we inherit the \textit{QPU} class and override the \textit{launchKernel} method. Within the method, we create a \textit{CircuitJobRequest} using the quantum kernel's corresponding \textit{Quake} representation.\
Hence, a quantum application can target remote and \gls{hpc} workflows without any modification.\ 
From the user's perspective, the \gls{mqss} backend behaves identically to any native CUDA Quantum backend — the \gls{mqss} Client layer remains entirely transparent.

\subsubsection{Python-based Qiskit and Pennylane Backend}\hfill\break
Qiskit Framework \cite{qiskit} provides abstractions to create custom backends.\
Abstractions such as \textit{BackendV2} and \textit{JobV1}, which are required components to execute a Qiskit circuit.

To connect Qiskit to \gls{mqss} Client, we inherit the \textit{BackendV2} class to orchestrate the job submission and receive information about the available quantum resources and \textit{JobV1} class to create the \textit{CircuitJobRequest} based on the OpenQASM \cite{cross2017openquantumassemblylanguage} representation of the Qiskit Circuit.

Similarly, a custom \gls{mqss} client further enables PennyLane \cite{bergholm2018pennylane} job submissions through a custom \gls{mqss} PennyLane adapter, allowing users to use native PennyLane features within hybrid workflows, such as quantum optimization via the parameter-shift method or gradient-based optimizers. The way to enable that is through the implementation of a PennyLane device \cite{pennylane_devices}, where execution occurs via \gls{mqss} client calls. Similar to the Qiskit adapter, the client can also fetch dynamic and static metadata, such as the device's coupling map and basis gates. 

\subsubsection{MQSS Benchmarking Framework}\hfill\break
Another front-end module that \gls{mqss} Client enables is the MQSS Benchmarking Framework \cite{mqss_benchmarking_framework}.\
MQSS Benchmarking Framework is a software package that offers high-level interfaces to benchmarking workflows and further separates them into hardware, software, application, and simulation benchmarks.\ 
It provides abstract classes for adapters and benchmarks, enabling end users to implement their own.\
The module also defines core features and implementations of these abstract classes, namely the \gls{mqss} adapters, and a few core benchmarks, such as randomized benchmarking \cite{knill2008randomized}, QAOA \cite{farhi2014quantum}, and Quantum Volume \cite{cross2019validating}.\
\gls{mqss} Client serves as the common entry point and the runtime module across all adapters.\
It also enables fetching hardware benchmark metrics, such as gate and readout fidelities, or runtime execution characteristics (compilation duration per job, quantum/classical time, etc.), without the need to couple individual adapters, as it forwards the required metrics directly to the framework.

%% file: sections/05_evaluation.tex
\section{Evaluation} \label{sec:exps-evaluation}

\begin{table*}[t!]
\centering
\resizebox{\textwidth}{!}{%
\begin{tabular}{llccccccccccc}
\toprule
& & & & \multicolumn{3}{c}{\textbf{AQT20}} 
    & \multicolumn{3}{c}{\textbf{QExa20}} 
    & \multicolumn{3}{c}{\textbf{MAQCS}} \\
\cmidrule(lr){5-7} \cmidrule(lr){8-10} \cmidrule(lr){11-13}
\textbf{Lang.} & \textbf{TE (s)} & \textbf{CC (ms)} & \textbf{RQ (ms)}
& \textbf{JSC (ms)} & \textbf{RES (s)} & \textbf{EQC (s)}
& \textbf{JSC (ms)} & \textbf{RES (s)} & \textbf{EQC (s)}
& \textbf{JSC (ms)} & \textbf{RES (s)} & \textbf{EQC (s)} \\
\midrule

\multicolumn{13}{l}{\textit{HPC Access Mode}} \\
\midrule

C++     & 28.55 & 5.25 & 9.98
& 10.64 & 2.23 & 13.85
& 8.62 & 1.90 & 4.13
& 13.24 & 1.86 & 6.04\\

Python  & 34.82 & 6.16 & 6.38 
& 6.06 & 1.87 & 20.23
& 9.11 & 1.83 & 4.21
& 6.22 & 2.11 & 3.92\\

\midrule
\multicolumn{13}{l}{\textit{Remote Access Mode}} \\
\midrule

C++  & 36.14 & 1.29 & 217.21
& 240.79 & 1.48 & 15.93
& 181.52 & 1.20 & 7.65
& 256.86 & 1.87 & 7.04 \\

Python  & 47.20 & 3.89 & 208.01
& 254.75 & 1.16 & 31.29
& 156.90 & 1.69 & 5.00
& 183.01 & 1.28 & 5.47\\

\bottomrule\\
\end{tabular}%
}
\caption{Results obtained from Validation Experiments.
TE: Total Execution, CC: Client Creation, RQ: Resource Query
JSC: Job Submission and Creation, RES: Result Retrieval.}
\label{tab:validation}
\end{table*}

\subsection{Evaluation Setup}

To evaluate the \gls{mqss} Client, we conducted a validation experiment and a performance experiment.\
The validation experiment aims to verify that the \gls{mqss} Client performs as expected in the real system.\
In addition, the performance experiment aims to evaluate the performance introduced by the \gls{mqss} Client.

\subsubsection{Validation Experiment}\hfill\break
For the validation experiment, we develop four standalone applications covering two programming languages (C++ and Python) and two access modes (\gls{hpc} and Remote), targeting the production \gls{mqss} server deployed at \gls{lrz}.

For the HPC access mode experiments, we used \gls{hpc} resources at the LRZ, which \gls{qc} systems are integrated into.\
The experiments are submitted to the resources using Slurm \cite{slurm}; then the quantum parts are offloaded to the quantum accelerator.\
The classical node has 256 GB of memory and features 2 Intel CPU sockets (Intel Xeon Platinum 8360Y @ 2.40 GHz), each with 36 physical cores.

For the remote access mode experiments, we use a Docker environment based on the \texttt{python:3.11-bookworm} image, running on a MacBook Pro with an Apple M3 Pro Chip and 18 GB of memory, to reflect typical remote user conditions.\

During these experiments, we target three different quantum resources: (1) 20-qubit Ion-Trap System from AQT (AQT20),  (2) 20-qubit superconducting system from IQM (QExa20), and (3) 16-qubit neutral atom demonstrator from planqc (MAQCS). 


\subsubsection{Performance Experiment}\hfill\break
For the performance experiment, we evaluate the performance introduced by the \gls{mqss} Client by comparing it against a direct implementation across two access modes — HPC and Remote — and two programming languages — C++ and Python.\
The direct implementation constructs and submits job requests manually.\
The Python experiment additionally validates the choice of C++ as the core implementation language, as the Python interface is built on top of the C++ core via pybind11 \cite{pybind11}; any overhead difference between the two languages reflects the cost of the binding layer.\

To ensure reproducibility and minimize measurement noise, experiments are conducted in the test environment rather than in the production environment, where the system's load may affect performance results.\
While for the remote access experiments, we send requests to a remote server that is hosted by \gls{lrz}, for the \gls{hpc} access experiments, the RabbitMQ daemon and the software stack are hosted locally on the same machine as the client.\
As a result, the reported \gls{hpc} overhead reflects the cost of the abstraction layer alone, independent of network conditions.\
We note that submitted jobs are not executed on any quantum backend.

Each experiment run consists of the following steps:
\begin{itemize}
    \item \textbf{\gls{te}:} End-to-end runtime of the application.
    \item \textbf{\gls{cc} :} Initialization of the \gls{mqss} Client (not applicable for direct implementations).
    \item \textbf{\gls{rq} :} Retrieval of available quantum resources.
    \item \textbf{\gls{jsc} :} Construction and submission of the quantum job.
    \item \textbf{\gls{eqc}:} Backend-reported execution time of the quantum circuit, returned as part of the \textit{JobResult}.
    \item \textbf{\gls{res}} Estimated time for result transfer, computed as the difference between the blocking result acquisition time and the \gls{eqc}, i.e., $\text{RES} = T_{\text{getJobResult}} - \text{EQC}$.
\end{itemize}

We repeated the performance experiment 100 times to ensure statistical reliability; however, we performed the validation experiment once, as it targets a production system subject to queue scheduling and hardware availability.\

\begin{table}[h]
\centering
\resizebox{\linewidth}{!}{%
\begin{tabular}{llcccccccccccc}
\toprule
& & & & & \multicolumn{3}{c}{\textbf{Test Device}}  \\
\cmidrule(lr){6-8} 
\textbf{Lang.} & \textbf{Impl.} & \textbf{TE (s)} & \textbf{CC (ms)} & \textbf{RQ (ms)}
& \textbf{JSC (ms)} & \textbf{RES (s)} & \textbf{EQC (s)}\\
\midrule

\multicolumn{8}{l}{\textit{HPC Access Mode}} \\
\midrule

C++    & Direct & 2.31 & -- & 19.70 & 13.05 & 1.24 & 0.95 \\

C++    & Client & 2.42 & 20.25 & 24.12 & 14.16 & 1.26 & 0.97 \\

Python & Direct & 2.71 & -- & 54.15 & 52.50 & 1.26 & 1.04\\

Python & Client & 2.62 & 19.41 & 23.82 & 15.18 & 1.31 & 0.98 \\

\midrule
\multicolumn{8}{l}{\textit{Remote Access Mode}} \\
\midrule
C++    & Direct & 24.44 & -- & 258.77 & 185.63 & 1.46 & 22.46\\

C++    & Client & 24.64 & 1.91 & 273.12 & 185.17 & 1.36 & 22.74 \\

Python & Direct & 25.28 & -- & 247.09 & 204.83 & 1.95 & 22.54 \\

Python & Client & 25.03 & 3.69 & 250.31 & 186.37 & 1.43 & 22.84 \\
\bottomrule\\
\end{tabular}%
}
\caption{Results obtained from Performance Experiments.}
\label{tab:performance}
\end{table}

\subsection{Evaluation Results}
In this section, we present the results of our two experiments and evaluate them. 

Table \ref{tab:validation} shows the results of our validation experiments.\
The results demonstrate that \gls{mqss} Client performs as expected across all programming languages and access modes.\
The \gls{hpc} access mode outperforms the remote access mode across the programming languages.\
Even though the results are expected, the \gls{te} time is not a suitable metric for evaluating the \gls{mqss} Client, as it is highly dependent on the execution of the quantum circuit.

The \gls{eqc} shows variable elapsed time across the target devices, based not onl on access mode and programming language.\
This directly affects the \gls{te}, with on average 84.5\% of the \gls{te} and a range of 81\% to 88\%. 

Even though the effect of the \gls{cc}, introduced by the \gls{mqss} Client, is negligible, taking only $1.12*10^{-2}$\% on average, it varies more across access modes than across programming languages.\
This can be attributed to differences in the underlying communication infrastructure, whereas the \gls{hpc} access mode is responsible for declaring the required messaging queues.\
However, after the initial communication is established, \gls{hpc} access mode significantly outperforms \gls{hpc} requirements on \glspl{rq} and \glspl{jsc}, which rely on communication. 

Table \ref{tab:performance} shows the results of our performance experiments.\
Unlike validation experiments, we did not target the production environment.\
For remote access mode experiments, we target the staging environment with a mock device server; likewise, for \gls{hpc} access mode experiments, we created a local test environment that exhibits the same behavior except for the execution of the quantum circuit. 

In the \gls{hpc} access mode, the experiments mostly show predicted results across the access mode, language, and implementation type.\
The direct implementation in C++ outperforms the other configurations, followed by client implementations in C++ and Python, and lastly, the direct implementation in Python.\

The \gls{te} difference between the C++ implementations is 0.21 seconds, and approximately 9\%.\
The main cause of the difference is the \gls{eqc}, which is 0.15 seconds longer in the client implementation.\
Since it is not directly caused by the \gls{mqss} Client itself, we can disregard it for more accurate evaluation.\
Another significant difference is the \gls{res}, which is 0.06 seconds higher in the client implementation.\
It may be caused by the state of the communication infrastructure, along with the overhead introduced by the \gls{mqss} Client.

The \gls{te} difference between the Python implementations is 0.9 seconds, and approximately 2\%, where the client implementation outperforms.\
The main differences observed on the \gls{res} and \gls{eqc}, where they differ by 0.5 and 0.6 seconds.\
While the direct implementation outperforms on \gls{res}, the client implementation outperforms on other parts, even if we disregard the \gls{eqc}.\

The \gls{te} difference between client implementations is 0.2 seconds and approximately 8\%. where the C++ implementation outperforms.\
The results show similar behaviors.\
However, we observe a noticeable difference in \gls{jsc} and \gls{res}.\
Since \texttt{JobRequest} and \texttt{JobResult} carry larger data, this situation may have made the overhead added by these Python bindings more noticeable.

We have drawn the following conclusions from our experiments:
\begin{enumerate}
    \item The overhead introduced by the \gls{mqss} Client is negligible.
    \item Using C++ in the core of the \gls{mqss} Client provides better performance.
    \item Python bindings cause a negligible overhead.
\end{enumerate}

Experiments for the remote access mode lead to the same conclusions.\
Only a significant difference on \gls{cc} is observed compared to \gls{hpc} access mode.\
As stated earlier, this can be explained by the underlying communication infrastructure as \gls{cc} in the \gls{hpc} mode responsible for declaring messaging queues.

%% file: sections/06_conclusion.tex
\section{Conclusion and Future Work} \label{sec:conclusion}

In this paper, we present \gls{mqss} Client, a unifying, context-aware access layer and programming library that decouples the programming model and the underlying compilation and runtime stack.\
\gls{mqss} Client addresses \gls{sdk} fragmentation in \gls{hpc} facilities, enabling diverse programming models to target the same quantum infrastructure without modification.\
Our evaluation shows that \gls{mqss} Client introduces negligible overhead compared to direct implementation in C++. Notably, the Python binding outperforms the direct Python implementation across both access modes, demonstrating the benefit of building the client on a compiled C++ core.\
As future work, we plan to extend the supported job types to include Annealing and Pulse-Level jobs, and to broaden compatibility with additional programming models beyond the current instantiations.\

%% file: sections/07_acknowledgments.tex
\section{Acknowledgments}
The authors used Grammarly, Claude, and ChatGPT for language editing, grammar check, and to improve the quality of the article.\
All content was reviewed and edited by the authors, who take full responsibility for the final work. \
This work is supported by the German Federal Ministry of Research, Technology, and Space (BMFTR) with the grants 13N15689 (DAQC), 13N16063 (Q-Exa), 13N16078 (MUNIQC-Atoms), 13N16187 (MUNIQC-SC), 13N16690 (Euro-Q-Exa), 13N16894 (MAQCS), European fundings 101114305 (Millenion), 10111394\-6 (OpenSuperQPlus), 101194491 (QEX), and the Bavarian State Ministry of Science and the Arts (StMWK) through funding, as part of Munich Quantum Valley, Q-DESSI.

%% file: bibtex.bib
@inproceedings{mqss-paper,
author = {Burgholzer, Lukas and Echavarria, Jorge and Hopf, Patrick and Stade, Yannick and Rovara, Damian and Schmid, Ludwig and Kaya, Erc\"{u}ment and Mete, Burak and Farooqi, Muhammad Nufail and Chung, Minh and De Pascale, Marco and Schulz, Laura and Schulz, Martin and Wille, Robert},
title = {The Munich Quantum Software Stack: Connecting End Users, Integrating Diverse Quantum Technologies, Accelerating HPC},
year = {2026},
isbn = {9798400720673},
publisher = {Association for Computing Machinery},
address = {New York, NY, USA},
url = {https://doi.org/10.1145/3773656.3773669},
doi = {10.1145/3773656.3773669},
booktitle = {Proceedings of the Supercomputing Asia and International Conference on High Performance Computing in Asia Pacific Region},
pages = {55–67},
numpages = {13},
location = {
},
series = {SCA/HPCAsia '26}
}

@INPROCEEDINGS{disaggregated-paper,
  author={Kaya, Ercüment and Echavarria, Jorge and Farooqi, Muhammad Nufail and Swierkowska, Aleksandra and Hopf, Patrick and Mete, Burak and Burgholzer, Lukas and Wille, Robert and Schulz, Laura and Schulz, Martin},
  booktitle={SC24-W: Workshops of the International Conference for High Performance Computing, Networking, Storage and Analysis}, 
  title={A Software Platform to Support Disaggregated Quantum Accelerators}, 
  year={2024},
  volume={},
  number={},
  pages={1646-1653},
  doi={10.1109/SCW63240.2024.00205}}

@INPROCEEDINGS{qpi,
  author={Kaya, Ercüment and Mete, Burak and Schulz, Laura and Farooqi, Muhammad Nufail and Echavarria, Jorge and Schulz, Martin},
  booktitle={2024 IEEE International Conference on Quantum Computing and Engineering (QCE)}, 
  title={QPI: A Programming Interface for Quantum Computers}, 
  year={2024},
  volume={02},
  number={},
  pages={286-291},
  doi={10.1109/QCE60285.2024.10293}}

@misc{pennylane,
      title={PennyLane: Automatic differentiation of hybrid quantum-classical computations}, 
      author={Ville Bergholm and Josh Izaac and Maria Schuld and Christian Gogolin and Shahnawaz Ahmed and Vishnu Ajith and M. Sohaib Alam and Guillermo Alonso-Linaje and B. AkashNarayanan and Ali Asadi and Juan Miguel Arrazola and Utkarsh Azad and Sam Banning and Carsten Blank and Thomas R Bromley and Benjamin A. Cordier and Jack Ceroni and Alain Delgado and Olivia Di Matteo and Amintor Dusko and Tanya Garg and Diego Guala and Anthony Hayes and Ryan Hill and Aroosa Ijaz and Theodor Isacsson and David Ittah and Soran Jahangiri and Prateek Jain and Edward Jiang and Ankit Khandelwal and Korbinian Kottmann and Robert A. Lang and Christina Lee and Thomas Loke and Angus Lowe and Keri McKiernan and Johannes Jakob Meyer and J. A. Montañez-Barrera and Romain Moyard and Zeyue Niu and Lee James O'Riordan and Steven Oud and Ashish Panigrahi and Chae-Yeun Park and Daniel Polatajko and Nicolás Quesada and Chase Roberts and Nahum Sá and Isidor Schoch and Borun Shi and Shuli Shu and Sukin Sim and Arshpreet Singh and Ingrid Strandberg and Jay Soni and Antal Száva and Slimane Thabet and Rodrigo A. Vargas-Hernández and Trevor Vincent and Nicola Vitucci and Maurice Weber and David Wierichs and Roeland Wiersema and Moritz Willmann and Vincent Wong and Shaoming Zhang and Nathan Killoran},
      year={2022},
      eprint={1811.04968},
      archivePrefix={arXiv},
      primaryClass={quant-ph},
      url={https://arxiv.org/abs/1811.04968}, 
}

@misc{qiskit,
      title={Quantum computing with {Q}iskit},
      author={Javadi-Abhari, Ali and Treinish, Matthew and Krsulich, Kevin and Wood, Christopher J. and Lishman, Jake and Gacon, Julien and Martiel, Simon and Nation, Paul D. and Bishop, Lev S. and Cross, Andrew W. and Johnson, Blake R. and Gambetta, Jay M.},
      year={2024},
      doi={10.48550/arXiv.2405.08810},
      eprint={2405.08810},
      archivePrefix={arXiv},
      primaryClass={quant-ph}
}

@mics{neven-2025,
	author = {Neven, Hartmut},
	month = {6},
	title = {{Meet Willow, our state-of-the-art quantum chip}},
	year = {2025},
	url = {https://blog.google/innovation-and-ai/technology/research/google-willow-quantum-chip/},
}

@misc{cross2017openquantumassemblylanguage,
      title={Open Quantum Assembly Language}, 
      author={Andrew W. Cross and Lev S. Bishop and John A. Smolin and Jay M. Gambetta},
      year={2017},
      eprint={1707.03429},
      archivePrefix={arXiv},
      primaryClass={quant-ph},
      url={https://arxiv.org/abs/1707.03429}, 
}

@manual{qir,
  title         = {{QIR Specification}},
  author        = {{QIR Alliance}},
  year          = {2021},
  url           = {https://github.com/qir-alliance/qir-spec},
  note          = {Also see \url{https://qir-alliance.org}}
}

@misc{mqss_interfaces,
  title        = {{MQSS Interfaces Documentation}},
  author       = {{Munich Quantum Software Stack (MQSS) Contributors}},
  year         = {2025},
  howpublished = {\url{https://munich-quantum-software-stack.github.io/MQSS-Interfaces/}},
  note         = {Accessed: \today}
}

@article{bacher2025quantum,
  title={Quantum resources in resource management systems},
  author={Bacher, Utz and Birmingham, Mark and Carothers, Christopher D and Damin, Andrew and Calaza, Carlos D Gonzalez and Karnad, Ashwin Kumar and Mensa, Stefano and Moreau, Matthieu and Nober, Aurelien and Ohtani, Munetaka and others},
  journal={arXiv preprint arXiv:2506.10052},
  year={2025}
}

@article{
  Ittah2024,
  doi = {10.21105/joss.06720},
  url = {https://doi.org/10.21105/joss.06720},
  year = {2024},
  publisher = {The Open Journal},
  volume = {9},
  number = {99},
  pages = {6720},
  author = {David Ittah and Ali Asadi and Erick Ochoa Lopez and Sergei Mironov and Samuel Banning and Romain Moyard and Mai Jacob Peng and Josh Izaac},
  title = {Catalyst: a Python JIT compiler for auto-differentiable hybrid quantum programs},
  journal = {Journal of Open Source Software}
}

@INPROCEEDINGS{cuda-q,
  author={Kim, Jin-Sung and McCaskey, Alex and Heim, Bettina and Modani, Manish and Stanwyck, Sam and Costa, Timothy},
  booktitle={2023 60th ACM/IEEE Design Automation Conference (DAC)}, 
  title={CUDA Quantum: The Platform for Integrated Quantum-Classical Computing}, 
  year={2023},
  volume={},
  number={},
  pages={1-4},
  doi={10.1109/DAC56929.2023.10247886}}

@misc{cq,
      title={Introducing CQ: A C-like API for Quantum Accelerated HPC}, 
      author={Oliver Thomson Brown and Mateusz Meller and James Richings},
      year={2025},
      eprint={2508.10854},
      archivePrefix={arXiv},
      primaryClass={cs.DC},
      url={https://arxiv.org/abs/2508.10854}, 
}

@inproceedings{qdmi,
    title = {{QDMI -- Quantum Device Management Interface: A Standardized Interface for Quantum Computing Platforms}},
    shorttitle = {{QDMI -- Quantum Device Management Interface}},
    booktitle = {IEEE International Conference on Quantum Computing and Engineering (QCE)},
    author = {Wille, Robert and Schmid, Ludwig and Stade, Yannick and Echavarria, Jorge and Schulz, Martin and Schulz, Laura and Burgholzer, Lukas},
    date = {2024},
}

@article{bergholm2018pennylane,
  title={Pennylane: Automatic differentiation of hybrid quantum-classical computations},
  author={Bergholm, Ville and Izaac, Josh and Schuld, Maria and Gogolin, Christian and Ahmed, Shahnawaz and Ajith, Vishnu and Alam, M Sohaib and Alonso-Linaje, Guillermo and AkashNarayanan, Bharath and Asadi, Ali and others},
  journal={arXiv preprint arXiv:1811.04968},
  year={2018}
}

@misc{mqss_benchmarking_framework,
  title        = {MQSS Benchmarking Framework},
  author       = {{Munich Quantum Software Stack (MQSS)}},
  year         = {2026},
  howpublished = {\url{https://github.com/Munich-Quantum-Software-Stack/MQSS-Benchmarking-Framework}},
  note         = {Accessed: 2026-04-01},
}

@article{knill2008randomized,
  title   = {Randomized benchmarking of quantum gates},
  author  = {Knill, Emanuel and Leibfried, Dietrich and Reichle, Reinhard and others},
  journal = {Physical Review A},
  volume  = {77},
  number  = {1},
  pages   = {012307},
  year    = {2008},
  doi     = {10.1103/PhysRevA.77.012307}
}

@article{farhi2014quantum,
  title   = {A Quantum Approximate Optimization Algorithm},
  author  = {Farhi, Edward and Goldstone, Jeffrey and Gutmann, Sam},
  journal = {arXiv preprint arXiv:1411.4028},
  year    = {2014},
  url     = {https://arxiv.org/abs/1411.4028}
}

@article{cross2019validating,
  title   = {Validating quantum computers using randomized model circuits},
  author  = {Cross, Andrew W. and Bishop, Lev S. and Sheldon, Sarah and Nation, Paul D. and Gambetta, Jay M.},
  journal = {Physical Review A},
  volume  = {100},
  number  = {3},
  pages   = {032328},
  year    = {2019},
  doi     = {10.1103/PhysRevA.100.032328}
}

@misc{curl,
  author       = {Stenberg, Daniel},
  title        = {{curl}},
  howpublished = {\url{https://curl.se}},
  year         = {1998},
  note         = {Accessed: 2026-04-06}
}

@software{Lohmann_JSON_for_Modern_2025,
author = {Lohmann, Niels},
license = {MIT},
month = apr,
title = {{JSON for Modern C++}},
url = {https://github.com/nlohmann},
version = {3.12.0},
year = {2025}
}

@misc{rabbitmqc,
  title = {rabbitmq-c: RabbitMQ C client library},
  author = {{rabbitmq-c developers}},
  year = {2024},
  howpublished = {\url{https://github.com/alanxz/rabbitmq-c}},
  note = {Accessed: 2026-04-06}
}

@misc{pybind11,
   author = {Wenzel Jakob and Jason Rhinelander and Dean Moldovan},
   year = {2017},
   note = {https://github.com/pybind/pybind11},
   title = {pybind11 -- Seamless operability between C++11 and Python}
}

@InProceedings{slurm,
author="Jette, Morris A.
and Wickberg, Tim",
editor="Klus{\'a}{\v{c}}ek, Dalibor
and Corbal{\'a}n, Julita
and Rodrigo, Gonzalo P.",
title="Architecture of the Slurm Workload Manager",
booktitle="Job Scheduling Strategies for Parallel Processing",
year="2023",
publisher="Springer Nature Switzerland",
address="Cham",
pages="3--23",
isbn="978-3-031-43943-8"
}

@misc{pennylane_devices,
  author       = {{Xanadu Quantum Technologies}},
  title        = {PennyLane Quantum Devices},
  year         = {2026},
  howpublished = {\url{https://pennylane.ai/devices}},
  note         = {Accessed: 2026-04-20}
}

@inproceedings{mansfield2025iqmfirstpractice,
    author = {Mansfield, Eric and Seegerer, Stefan and Vesanen, Panu and Echavarria, Jorge and Farooqi, Muhammad Nufail and Mete, Burak and Schulz, Laura},
    title = {First Practical Experiences Integrating Quantum Computers with HPC Resources: A Case Study With a 20-qubit Superconducting Quantum Computer},
    year = {2025},
    isbn = {9798400718717},
    publisher = {Association for Computing Machinery},
    address = {New York, NY, USA},
    url = {https://doi.org/10.1145/3731599.3767551},
    doi = {10.1145/3731599.3767551},
    booktitle = {Proceedings of the SC '25 Workshops of the International Conference for High Performance Computing, Networking, Storage and Analysis},
    pages = {1842–1850},
    numpages = {9},
    location = {},
    series = {SC Workshops '25}
}
